\documentclass{IEEEcsmag}

\usepackage[colorlinks,urlcolor=blue,linkcolor=blue,citecolor=blue]{hyperref}
\expandafter\def\expandafter\UrlBreaks\expandafter{\UrlBreaks\do\/\do\*\do\-\do\~\do\'\do\"\do\-}
\usepackage{upmath,color}

\usepackage{graphicx}
\usepackage{caption}
\usepackage{subcaption}

\usepackage[superscript,biblabel]{cite} 

\usepackage{listings}

\usepackage[colorinlistoftodos,prependcaption]{todonotes}

\jvol{XX} 
\jnum{XX}
\paper{8}
\jmonth{Month}
\jname{IEEE Software}
\jtitle{Generative AI for Software Engineering}
\pubyear{2024}
\begin{document}

% et al
\bstctlcite{IEEEexample:BSTcontrol}

\sptitle{Generative AI for Software Engineering}

%\title{Test-driven Assessment and Enhancement of Generative AI Reliability}

\title{N-Version Assessment and Enhancement of Generative AI}

\author{Marcus Kessel}
\affil{University of Mannheim, Mannheim, 68159, Germany}

\author{Colin Atkinson}
\affil{University of Mannheim, Mannheim, 68159, Germany}

\markboth{Generative AI for Software Engineering}{Generative AI for Software Engineering}

% max: 150 words
% currently ~ 136words (11.09.24 - 11:34)
\begin{abstract}\looseness-1
Generative AI (GAI) holds great potential to improve software engineering productivity, but its untrustworthy outputs, particularly in code synthesis, pose significant challenges. The need for extensive verification and validation (V\&V) of GAI-generated artifacts may undermine the potential productivity gains. This paper proposes a way of mitigating these risks by exploiting GAI's ability to generate multiple versions of code and tests to facilitate comparative analysis across versions. Rather than relying on the quality of a single test or code module, this ``differential GAI'' (D-GAI) approach promotes more reliable quality evaluation through version diversity. We introduce the Large-Scale Software Observatorium (LASSO), a platform that supports D-GAI by executing and analyzing large sets of code versions and tests. We discuss how LASSO enables rigorous evaluation of GAI-generated artifacts and propose its application in both software development and GAI research.
\end{abstract}

\maketitle

\chapteri{B}y automatically generating software artifacts, including code, from small, natural language prompts, generative AI (GAI) has the potential to dramatically boost the productivity of software engineers. However, while the creativity of GAI systems is essentially boundless, the ``quality'' of the artifacts they generate is not, significantly limiting their trustworthiness in engineering domains where low quality is costly. Effectively judging, and if necessary enhancing, the quality of GAI-generated artifacts will therefore be critical to its success in practical software engineering (SE).

Software quality can be defined by various properties, but measuring some of the most critical quality attributes -- specifically semantic (i.e., behavioral) properties -- remains undecidable in most cases (cf. Rice's theorem\cite{riceTheorem}). As a result, it is impossible, in general, to develop a tool that can automatically verify non-trivial semantic properties of arbitrary code modules, including their functional correctness relative to specifications or previous versions. This limitation is a key reason why modern software projects often allocate over 50\% of their total effort to verification and validation (V\&V) techniques such as testing \cite{ammann2016introduction}. Consequently, any potential productivity gains from using GAI could be significantly diminished -- or even negated -- by the additional V\&V work required to mitigate its inherent unreliability.

At first glance, it might seem feasible to address the V\&V challenges posed by GAI by using it to automatically generate additional V\&V artifacts, such as tests. However, these artifacts -- like all GAI-generated outputs -- are also of uncertain quality. This creates a dilemma: verifying code modules of uncertain quality with verification artifacts (e.g., tests) that are equally unreliable. The only practical way to break this cycle and achieve genuine confidence in the quality of GAI-generated code, without resorting to excessive manual effort, is to generate multiple, diverse versions of both code modules and tests \cite{10.1145/2807593} and conduct extensive comparative analysis of their executions.

The concept of comparing multiple executions of similar versions of code modules has a long history across various branches of SE. It appears in approaches such as differential testing for software V\&V \cite{8804465}, N-version programming for enhancing software reliability \cite{1701972}, mutation testing for assessing test quality, and the use of ``cross-checking'' oracles in automated test generation \cite{6963470}. While these methods differ in their objectives and the stages of a software artifact's lifecycle in which they are applied, they all share a common core: comparing the executions of multiple tests on multiple versions of code modules expected to deliver specific functionality. This comparative approach is also valuable in experimentally evaluating tools used to create or modify these versions, including tools for code refactoring, program repair, code recommendation, and GAI for code generation.

The premise of this paper is that the future success of GAI in mainstream SE projects will be significantly enhanced by large-scale, ``comparative'' or ``differential'' testing of GAI-generated artifacts. However, current testing frameworks are not suited for this purpose, as they are designed to test individual code modules. Even existing differential testing tools offer limited user feedback on discrepancies. Moreover, these tools lack the decision-making capabilities and data-driven insights of code recommenders, which take into account users' weighted preferences for functional and non-functional properties to select the best code modules or tests based on comprehensive criteria \cite{5235134}.

In this paper, we explain how the ``differential GAI'' (D-GAI) approach to code and test synthesis could significantly accelerate GAI-driven SE in both development and research. To illustrate this, we introduce the Large-Scale Software Observatorium (\textsc{LASSO}), a new software observation platform designed to support the D-GAI approach. \textsc{LASSO} provides domain-specific languages, data structures \cite{kesselOpenScience24}, and an execution ``arena'' that facilitates the execution of large numbers of tests across numerous code module versions. By analyzing the resulting observational data, functional and non-functional differences and patterns can be identified, leading to improved code quality and reliability -- regardless of how the versions or tests were generated.

\textsc{LASSO} can also be used to conduct large-scale, reproducible studies on the effectiveness of GAI models, with the observational data serving as valuable training input to further enhance GAI systems. After presenting a D-GAI scenario based on a typical code generation task, we outline how a D-GAI engine could be realized using a platform like \textsc{LASSO}. Finally, we explore how D-GAI can improve the utility of GAI in practical SE projects while also providing a means to assess and enhance GAI systems themselves.

\vspace*{-8pt}
\section{EXAMPLE D-GAI SCENARIO}
\label{sec:gai_scenario}

As an illustration of a D-GAI scenario, suppose a team of software developers wants to use one of the latest GAI code generation models to help accelerate their project. Some of the most well-known code models currently available  include GPT-3.5/-4 Turbo, Codex (GitHub CoPilot), CodeGen and InCoder (see \cite{liu2023is}), which are all based on the Transformer-driven LLM technology popularized by ChatGPT. These can perform a variety of SE tasks, but one of the most canonical is synthesizing code modules from natural language descriptions of the desired functionality. In the GAI community this is usually referred to as a code completion ``problem'' \cite{multiple2023}, but essentially provides a service called ``code recommendation'' in the traditional software reuse community \cite{5235134}.

To get such a code generation model (i.e., code model for short) to recommend a code module, developers have to provide it with a ``prompt`` that describes the desired software functionality in some way. For example, the following prompt is for a method that calculates the greatest common divisor (GCD) of two integers. After identifying the signature of the method (here in Python syntax), the prompt provides a short natural language description of the desired functionality as well as two tests a correct implementation is expected to pass.

\begin{lstlisting}[language=Python]
def gcd(a: int, b: int) -> int:
  """ Return a greatest common divisor
  of two integers a and b
  >>> gcd(3, 7)
  1
  >>> gcd(10, 15)
  5"""
  (body to be generated)
\end{lstlisting}

On receiving such a prompt, a code model returns a recommended implementation of the desired functionality. State-of-the-art code models are able to provide correct recommendations of this size a surprisingly high proportion of times. Recent experiments (e.g., MultiPL-E \cite{multiple2023}, EvalPlus \cite{liu2023is}) and leaderboards\footnote{\url{https://evalplus.github.io/leaderboard.html}} have achieved significant success rates on popular benchmarks, with some exceeding 80\%. However, while these rates are high from a GAI research point of view, they are not particularly high for practical software development (cf. \cite{10174227}). Thus, in order to be able to actually use a recommended code module in a real project, developers would probably need to test it extensively.

In the D-GAI version of this scenario, developers would receive the same basic service, driven by the same prompt, but the ``quality'' of the recommended code module would usually be much higher. This is achieved by means of a special tool or platform (i.e., a D-GAI engine) that automatically performs the code recommendation request multiple times, possibly using multiple code models, to generate N versions of the desired code module, and then automatically compares them using multiple, automatically generated tests. The D-GAI engine then returns the best version based on specific ranking criteria. 
%All this work is performed ``under the hood'', so that users are unaware of how the final results are generated. 
Applying such an N-version approach has numerous important benefits ---

\begin{itemize}
	\item[{\ieeeguilsinglright}] {\it Results Aggregation} --- the D-GAI engine is able to aggregate the results of many GAI code recommendation requests, from one or more code models. Even if one code model is used, this is advantageous because they are inherently non-deterministic in their behavior, and thus can give different results for the same prompt. When more code models are available, obviously the best results can be selected. Since it aggregates results, a D-GAI engine will always deliver a result that is at least as good as a single-use realization of the code recommendation service.
	
	\vspace*{3pt}

	\item[{\ieeeguilsinglright}] {\it Semantic Awareness} --- A major benefit of the D-GAI approach is that by actually executing all multiple candidates (i.e., N versions), a result can be returned that genuinely passes the required tests. The users of the recommendation service can then have confidence that the returned result, if there is one, truly has the behavior demanded by the tests. This not only increases confidence in the results, but obviates the need for further testing with the prompt tests. Code models themselves cannot give a guarantee that the recommended code passes the prompt tests, because they are unaware of the true behavior of software code \cite{kessel2024morescient}. This idea corresponds to the notion of test-driven search in the software reuse community \cite{kesselNextGen2024}.

	\vspace*{3pt}

	\item[{\ieeeguilsinglright}] {\it Observational Measurement} --- Another major benefit of executing the candidates is that their comparison can include dynamic code metrics, such as performance and coverage, as well as static code metrics (e.g., measurements obtained by applying GQM  \cite{basili1994goal}). Code models are unable to consider the former because they are unaware of the true execution properties of the code modules they generate. They also do not apply classic static metrics in the traditional sense, which a D-GAI engine could easily do.

\end{itemize}\vspace*{3pt}

Of course, the extra testing and analysis involved in D-GAI comes at the cost of significantly lower response times. However, considering the widespread adoption of test-driven development and continuous integration practices in modern SE, we envision GAI code recommendations being primarily used in an asynchronous, offline mode that will not disrupt developers' workflows. As such, D-GAI will likely become a natural extension of the offline capabilities offered by powerful continuous integration platforms, further streamlining the development process. Moreover, even for direct use, with a powerful D-GAI engine, the code recommendation response times are reasonable, depending on the number of module versions explored. In a recent experiment \cite{kesselNextGen2024}, average response times for $1.000$ code modules harvested from code repositories were \textasciitilde 4 minutes on modern hardware.

While employing code models may introduce costs due to additional token generation from multiple models or repeated iterations on a single model to obtain N versions of the code modules and tests, the D-GAI approach offers a rapid feedback loop reminiscent of agile methodologies. This helps developers to making informed, data-driven decisions, avoiding the pitfalls of blindly integrating generated code. By minimizing technical debt and potential risks, this approach can lead to significant long-term cost savings through reduced rework, debugging, and maintenance.

\vspace*{-8pt}
\section{TOWARDS A D-GAI ENGINE}
\label{sec:dgai_engine}

In this section, we delve deeper into the architecture of a D-GAI engine and outline key components necessary for its implementation. To facilitate the discussion, we employ our \textsc{LASSO} platform\footnote{\url{https://softwareobservatorium.github.io/}}, which was designed specifically to support the N-version comparison layer required to realize D-GAI on top of standard code models. While \textsc{LASSO} is not exclusively focused on D-GAI, it provides a suitable foundation for implementing the D-GAI approach. As well as a powerful testing platform and large code repository, \textsc{LASSO} offers several dedicated languages and data structures for N-Version comparison.

\begin{figure*}%[h]
	\centerline{\includegraphics[width=\textwidth,scale=1.0,trim=0cm 3cm 6cm 0cm,clip]{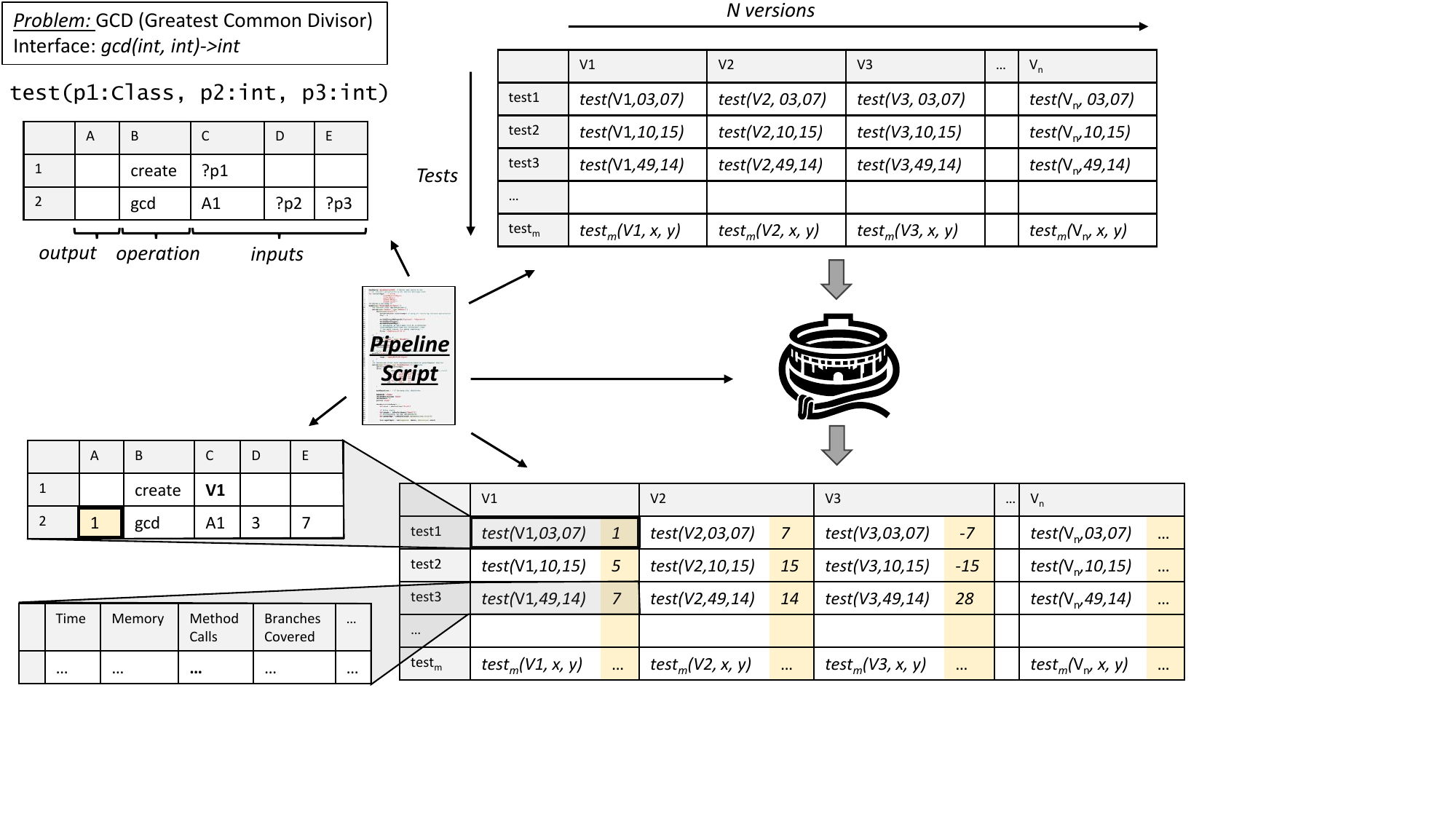}}
	\caption{Towards a D-GAI Engine -- Data Structures and Execution Arena for N-version Analysis}\vspace*{-5pt}
	\label{fig:data_structures}
\end{figure*}

Figure \ref{fig:data_structures} illustrates how the GCD code recommendation scenario described earlier can be efficiently implemented using \textsc{LASSO}. The top left section of the diagram introduces \textsc{LASSO}'s dedicated ``sequence sheet'' notation, used to store stimulus-response pairs for sequences of method invocations. A sequence sheet functions similarly to a method with a corresponding signature, but instead of using traditional code for its body, it uses a table format. Each row in this table represents an individual method invocation, while the columns represent different components: column \textit{B} designates the called method, columns \textit{C} to \textit{E} hold the input parameters, and column \textit{A} captures the output. Users can either represent their test components directly as sequence sheets or have them generated automatically from the traditional prompt format shown earlier.

On receiving a code recommendation prompt, \textsc{LASSO} can generate a so-called stimulus matrix (SM) of the form shown on the top right-hand side of Figure \ref{fig:data_structures} and populate it with multiple implementations of the desired functionality (as columns) and multiple tests of the desired functionality (as rows). The N versions can be obtained from a variety of sources, including from \textsc{LASSO}'s own code corpus, but in the D-GAI scenario they would be generated by multiple code generation requests to one or more code models. If only one code model is available, N versions can be obtained either (a) by invoking it many times since the code synthesis behavior is non-deterministic, or (b) by varying the various model parameter choices like the temperature parameter to fine-tune the level of generation behavior. To populate the SM with tests, any tests provided in the prompt are converted into sequence sheets and invocations, as shown by the first two rows in the example. More tests can also be added automatically through code models using generated prompts, or by applying standard unit test generation tools like \textsc{EvoSuite}\cite{10.1145/2685612} or \textsc{Randoop}.

Once the SM has been populated it can be input to \textsc{LASSO}'s dedicated test engine, the so-called ``arena'', where all the tests are executed on all N versions of the desired functionality. To cope with the high volume of executions that can be involved, the arena is implemented as a parallel, distributed architecture with load balancing and sandboxed (controllable and secure) code execution environments based on containerization. Once all the testing has been performed, the results are output as a stimulus response matrix (SRM), which has the same shape as the input SM, but contains all the runtime data observed during the execution process (i.e., outputs, execution time, methods called, branches covered etc.). As shown in the bottom right-hand side of Figure \ref{fig:data_structures}, the core advantage of using sequence sheets to define tests is that it allows SRMs to store the responses of the different module versions within the sequence sheets that provoked them. For example, the sequence sheet at the bottom left of the diagram shows that the invocation of GCD module version \textit{V1} with the values $3$ and $7$ (the first test in the prompt) returned the value $1$. SRMs therefore provide a persistent, analyzable record of the behavior of each version in response to each test.

One other important \textsc{LASSO} feature represented in Figure \ref{fig:data_structures} is the pipeline script used to drive the work flow steps involved in a D-GAI code recommendation request. This allows engineers to write new D-GAI services in a dedicated DSL.

Finally, the \textsc{LASSO} platform has a large underlying repository of executable software code, harvested from large open source code repositories, that can be accessed using classic code search technologies\cite{kesselNextGen2024}. Although the use of this repository is not essential for D-GAI, including human-written modules and tests harvested from the repository can enhance diversity, and thus improve the quality and trustworthiness of the overall results.

\vspace*{-8pt}
\section{D-GAI USE CASES \& SERVICES}
\label{sec:use_cases}

The \textit{LASSO} data structures described above are not essential for building a D-GAI engine. However, using them offers several advantages. Firstly, treating tests as ``first-class'' citizens in the differential comparison process allows them to be assessed and enhanced using D-GAI alongside the N module versions. Secondly, storing and analyzing execution observations offline with analysis tools like Python/Pandas, enables researchers and practitioners to assess, benchmark, and enhance the underlying code models across multiple coding problems.

\begin{figure*}[h]
	\centerline{\includegraphics[width=\textwidth,scale=1.0,trim=0cm 4cm 0cm 0cm,clip]{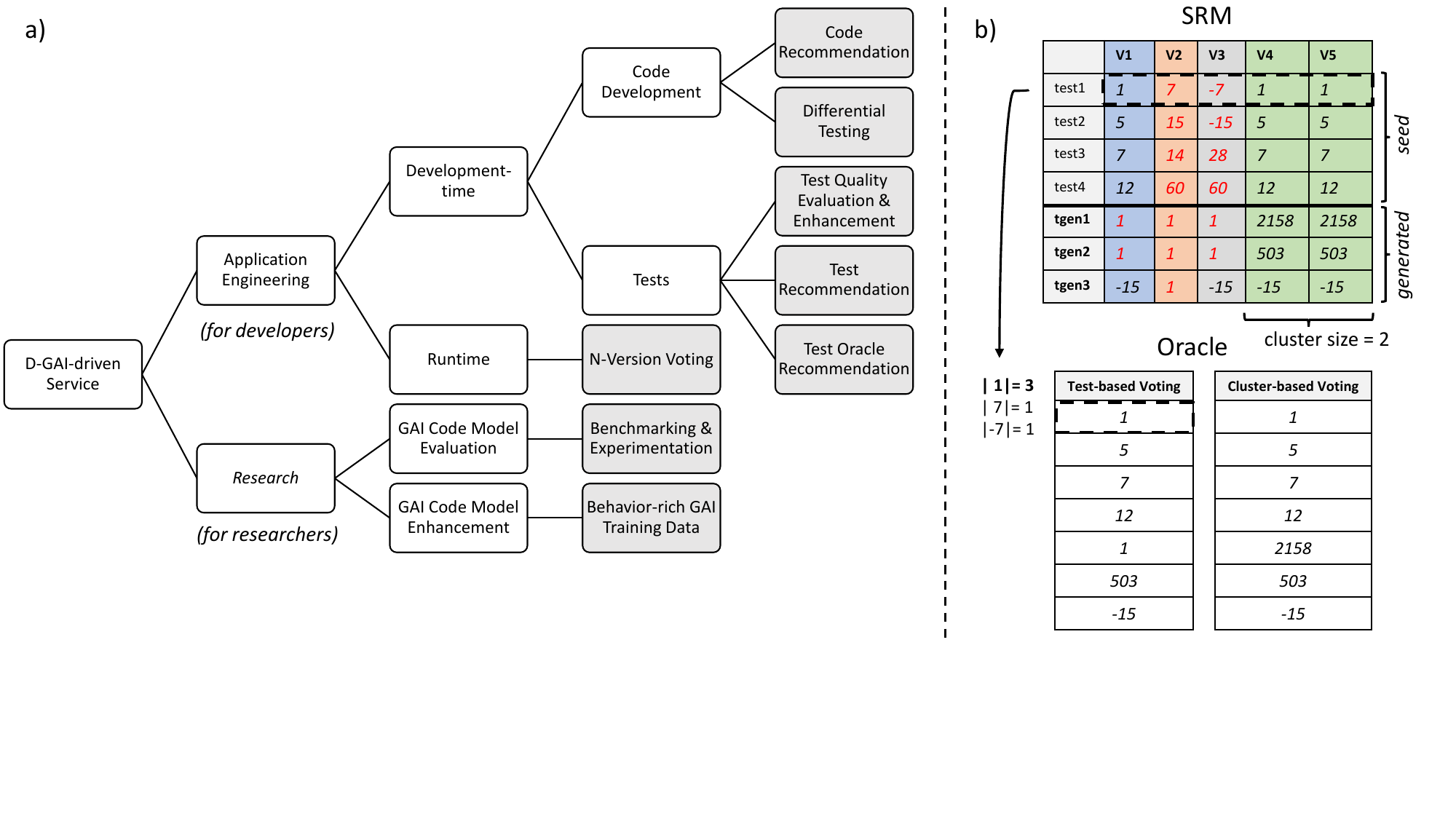}}
	\caption{(a) Differential GAI Use Cases and Services, and, (b) GCD Oracle  Example}\vspace*{-5pt}
	\label{fig:taxonomy}
\end{figure*}

Figure \ref{fig:taxonomy}(a) presents a comprehensive overview of the range of use cases and services that could be supported by a D-GAI engine of the kind outlined previously. It categorizes the services into three main tiers. The root-level distinction is between services aimed at assisting developers in building concrete software applications and those intended to aid researchers in evaluating or enhancing code models. Application Engineering services are further divided into Development-time and Runtime services. The latter focus on improving the quality of deployed applications during execution time as in classic N-version programming \cite{1701972}. Research services comprise GAI code model evaluation services for benchmarking and comparing different code models, as well as code model enhancement services for training and refining GAI models themselves. Within the Development-time application engineering category, subservices help developers to evaluate, create and enhance code and tests. In the remainder of this section, we will delve deeper into these leaf cases.

\vspace*{-8pt}

\subsection{Code Recommendation}

Code recommendation is the use case exemplified in the second section, and is one of the core use cases of D-GAI for software development. The basic idea resembles the test-driven search technology provided by some code search engines \cite{kesselNextGen2024}, and offers the same important benefit of improving the user's confidence in the correctness of the recommended module (or modules), since it has been verified to pass the tests supplied in the prompt. Code models themselves do not give a guarantee that the recommended code passes the prompt tests, because they are unaware of the true behavior of software code \cite{kessel2024morescient}. The basic difference is that whereas code search engines retrieve human-written code modules from code repositories, a D-GAI code recommendation service selects the most suitable code module from a set of GAI-synthesized modules. Executing the candidates also allows the final ranking of the candidates to include both dynamic metrics and static metrics. Code models can enhance code recommendations services not only by synthesizing large numbers of alternative module versions, but also by synthesizing additional tests.

\vspace*{-8pt}

\subsection{Differential Testing}

Differential testing is a widely practiced testing technique in modern SE projects which aims to improve the quality of a particular code module under development (i.e., ``base version'') by comparing its behavior to that of one or more alternative versions to identify discrepancies that may reveal faults or functional differences \cite{8804465}. It is most commonly applied in the form of regression testing where the alternative versions are older versions of the code module. This is because obtaining a diverse set of alternative versions of the base version (i.e., that are not simply clones) is a challenging task.

\vspace*{-8pt}

\subsection{Test Quality Evaluation \& Enhancement}

Test quality evaluation is perhaps the simplest of the test-oriented D-GAI services in which the prompt includes a set of tests and a base code module of the desired functionality, and the output is one or more test quality metrics. Code models are used only to generate new versions of the base version, which play the same role as the mutations in mutation testing \cite{ammann2016introduction}. After executing all the tests on all the versions, the D-GAI engine can calculate a version kill score, analogous to a mutation score to capture the quality of the supplied test sets.

Test enhancement services take this one step further by not only determining the version kill score but also leveraging this information to produce a minimal subset of input tests (i.e., test set minimization) that achieves the same (or a comparable) level of test effectiveness.

\vspace*{-8pt}

\subsection{Test Recommendation}

This service resembles the code recommendation service, but recommends new tests, or more precisely test sets. Here the prompt would include the base module the developer wishes to have tests for, rather than the tests the developer wishes to have a code module for. As in the previous cases, the code model would be asked to synthesize new tests as well as new versions of the supplied code whose execution outputs would then be compared for differences. The new versions would be used in the same way as in the previous two services to identify effective sets and return the minimum number needed to kill the largest possible number of versions.

\vspace*{-8pt}

\subsection{Test Oracle Recommendation}

Obtaining oracles to define correct behavior has traditionally been a big obstacle to test automation, and is still typically performed by hand \cite{6963470}. D-GAI can help by facilitating the creation of multiple ``verdicts'' on correct output values from N versions of the code modules. There are two basic strategies for doing this, illustrated at the bottom of Figure \ref{fig:taxonomy}(b) for the GCD example. The first approach, \textit{test-based voting}, identifies the oracle values by selecting the most common result for each test case across all N versions, while the second approach, \textit{cluster-based voting}, identifies the oracle values based on behavioral clustering over common outputs over all the tests. The former is essentially the approach taken in the original N-version programming approach \cite{1701972}. However, in the example in Figure \ref{fig:taxonomy}(b), test-based clustering is inconclusive, since none of the N versions match the oracle for all outputs, thus no variant is functionally correct and deemed to be of the highest quality according to these criteria. In contrast, the clustering-based approach leads to two versions being judged to as functionally correct, and thus of the highest quality -- namely \textit{V4} and \textit{V5} which are in the large (green) cluster. In the figure, outputs that deviate from the cluster-based voting oracle are shown in red font.

\subsection{Research}

To evaluate GAI-synthesized code artifact quality, researchers conduct experiments comparing generated artifacts in various coding problems. D-GAI facilitates this by aggregating SRMs for multiple problems and offering oracle-based quality assessments (cf. \cite{kesselOpenScience24}). As the number of code modules and tests grows, results generalizability and statistical power improve, allowing more precise reliability estimates. Additionally, combining D-GAI services mirrors existing benchmark enhancement approaches like \textsc{EvalPlus} \cite{liu2023is}. Furthermore, SRMs offer valuable observational data that can be used to train and improve code models, leading to better performance of GAI-synthesized code modules and tests.

\vspace*{-8pt}
\section{CONCLUSION}
\label{sec:conclusion}

This paper has introduced Differential GAI (D-GAI), a practical approach to improving the quality of generative AI (GAI)-produced software artifacts through the comparative analysis of multiple versions. By incorporating ``true'' semantic information derived from the execution of multiple versions, D-GAI enables more reliable quality assessments without depending on the trustworthiness of any single artifact.

Our key contribution is the introduction of the Large-Scale Software Observatorium (\textsc{LASSO}), a platform designed to support large-scale execution and analysis of code versions and tests. \textsc{LASSO} offers developers and researchers a powerful tool for assessing GAI-generated artifacts. By comparing different executions, LASSO facilitates the identification of behavioral patterns, functional discrepancies, and non-functional variations across versions, thereby enhancing the reliability and trustworthiness of GAI outputs.

\vspace*{-8pt}

%% END

\section{ACKNOWLEDGEMENT}

This research is funded by the Ministry of Science, Research and Arts Baden-Württemberg, Research Seed Capital (RiSC)

\def\refname{REFERENCES}

\bibliographystyle{myIEEEtran}
\bibliography{IEEEabrv,literature}

\begin{IEEEbiography}{Marcus Kessel}{\,} is a (postdoc) researcher in software engineering at the University of Mannheim, Germany. His research focuses on developing scalable and behavior-aware methods for software code analysis, integrating insights from static and dynamic program analysis with data-driven approaches. The author received the Ph.D. degree in computer science from University of Mannheim. Contact him at marcus.kessel@uni-mannheim.de.\vspace*{8pt}
\end{IEEEbiography}

% @Colin needs to be updated
\begin{IEEEbiography}{Colin Atkinson}{\,} is the leader of the software engineering group at the University of Mannheim, Germany. His current research interests include multi-level deep modeling, view-driven software engineering and scalable software analysis and observation. The author received the Ph.D. degree in computer science from Imperial College, London, and is currently a professor at the University of Mannheim. Contact him at colin.atkinson@uni-mannheim.de.\vspace*{8pt}
\end{IEEEbiography}

\end{document}